\documentclass[lettersize,journal]{IEEEtran}

\IEEEoverridecommandlockouts
\usepackage{cite}
\usepackage{amsmath,amssymb,amsfonts}
\usepackage{graphicx}
\usepackage{subfigure}
\usepackage{textcomp}
\usepackage{xcolor}
\usepackage[english]{babel}
\usepackage{amsthm}

\usepackage{booktabs}
\theoremstyle{definition}

\usepackage{subcaption}
\usepackage{multirow}
\usepackage[normalem]{ulem}  
\usepackage{tabularx}
\usepackage{amsmath,amsfonts}
\usepackage{array}
\usepackage[caption=false,font=normalsize,labelfont=sf,textfont=sf]{subfig}
\usepackage{textcomp}
\usepackage{stfloats}
\usepackage{url}
\usepackage{verbatim}
\usepackage{graphicx}
\usepackage{cite}
\begin{document}

\title{Fine-Grained Network Traffic Classification with Contextual QoS Profiling}

\author{Huiwen Zhang,~\IEEEmembership{Graduate Student Member,~IEEE,} Feng Ye,~\IEEEmembership{Senior Member,~IEEE}
\thanks{This project is partially supported by the U.S. National Science Foundation under Grant 2344341.}
\thanks{Huiwen and Feng Ye (corresponding) are with the Department of Electrical and Computer Engineering, University of Wisconsin-Madison, Wisconsin, WI, USA. Emails: \{hzhang2279, feng.ye\}@wisc.edu. 
}
}
\maketitle

\begin{abstract}
Accurate network traffic classification is vital for managing modern applications with strict Quality of Service (QoS) demands, such as edge computing, real-time XR, and autonomous systems. While recent advances in application-level classification show high accuracy, they often miss fine-grained in-app QoS variations critical for service differentiation. This paper proposes a hierarchical graph neural network (GNN) framework that combines a three-level graph representation with an automated QoS-aware assignment algorithm. The model captures multi-scale temporal patterns via packet aggregation, time-window clustering, and session-level behavior modeling. QoS priorities are derived using five key metrics (bandwidth, jitter, packet stability, burst frequency, and burst stability), processed through logarithmic transformation and weighted ranking. Evaluations across 14 usage scenarios from YouTube, Prime Video, TikTok, and Zoom show that the proposed GNN significantly outperforms state-of-the-art methods in service-level classification. The QoS-aware assignment further refines classification to enhance user experience. This work advances QoS-aware traffic classification by enabling precise in-app usage differentiation and adaptive service prioritization in dynamic network environments.

\end{abstract}

\section{Introduction}

Accurate network traffic classification (NTC) is essential for effective network management, particularly in the context of emerging and future applications that demand stringent performance guarantees. For example, applications such as edge cloud computing~\cite{5280678}, real-time extended reality~\cite{10.1145/3652595}, autonomous vehicle communication~\cite{8246845}, and industrial IoT~\cite{8819994} rely heavily on low-latency, high-throughput, and highly reliable network services. These applications introduce complex traffic patterns characterized by variable data rates, strict latency constraints, and dynamic resource demands that fluctuate in real time. As a result, precise NTC and Quality of Service (QoS) management have become increasingly critical to ensure application performance and user experience.

NTC has evolved significantly over the years, transitioning from traditional port-based methods and deep packet inspection to more advanced statistical and AI-driven techniques~\cite{10.1145/3457904,9791420, 9566310}. Recent developments in application-level NTC have achieved notable success, particularly in encrypted traffic classification, where models can infer application types without accessing payload content~\cite{AZAB2024676}. These approaches have demonstrated high accuracy in Internet and mobile application identification~\cite{9395707, 10.1145/3485832.3485925,9979671,9841019}, and have also significantly advanced anomaly detection~\cite{DUAN2023206,MA2021102215,pranto2022performance}, enabling proactive service assurance and threat mitigation. However, most existing methods focus on identifying the application or traffic type rather than capturing nuanced in-app QoS differences, such as distinguishing between video streaming at different resolutions or between interactive and background data flows. This limitation stems from their original design goals, which prioritized coarse-grained classification over the fine-grained service differentiation required for advanced QoS provisioning. To address these limitations, QoS-oriented NTC methods have emerged, aiming to classify traffic based on QoS attributes such as throughput, latency, and jitter~\cite{YU20181209,10459131}. While these methods provide valuable insights into network performance, they often rely on handcrafted features and manually defined service categories, which limit their scalability and adaptability to new or evolving applications. Furthermore, the rigid mapping between traffic patterns and QoS labels can hinder generalization across diverse network environments.

In this work, a new hierarchical graph neural network (GNN) framework is proposed to address these challenges. The proposed framework integrates a three-level hierarchical graph representation with an automated, magnitude-based QoS awareness assignment algorithm. It captures multi-scale temporal patterns through packet aggregation at Level-1, time window clustering at Level-2, and session-level behavioral modeling at Level-3. Classification is performed at the time window level (Level-2), leveraging both fine-grained packet-level features and broader session-level context to enhance usage pattern discrimination. 
Moreover, the newly developed QoS awareness assignment algorithm takes into consideration five different QoS attributes, including bandwidth, jitter, packet stability, burst frequency, and burst stability. By taking a logarithmic transformation of the raw values, each traffic flow can be dynamically assigned to a QoS class defined by all five metrics. A weighted ranking algorithm is further implemented to establish data-driven service priorities that are automatically adapting to traffic distribution characteristics. 
Evaluations are conducted on traffic traces collected from 14 different usage scenarios across YouTube, Prime Video, TikTok, and Zoom. The results demonstrate that the newly developed QoS-aware NTC enables fine-grained differentiation of in-app usage patterns (e.g., TikTok browsing vs. live streaming vs. long-form video) while ensuring appropriate QoS provisioning by prioritizing service quality preservation over resource optimization, comparing to a standard application-level NTC approach. 

The contributions of this work are fourfold. First, a novel three-level hierarchical graph representation is introduced, capturing temporal dependencies from packet-level interactions to session-level behaviors, thereby enabling fine-grained traffic classification beyond traditional application-level identification. Second, an automated magnitude-based QoS awareness assignment algorithm is developed, using logarithmic transformation and automated grouping to establish consistent, data-driven QoS priorities across diverse network conditions. Third, a QoS-aware training framework is proposed, incorporating composite loss functions and inference strategies that prioritize service quality preservation, ensuring over-provisioning rather than under-provisioning for critical applications. Finally, comprehensive experimental validation is conducted, demonstrating significant improvements in QoS Experience while maintaining competitive classification performance across 14 distinct usage scenarios spanning YouTube, Prime Video, TikTok, and Zoom.

The remainder of this paper is organized as follows: Section~\ref{sec:related_work} reviews related work in network traffic classification and QoS-aware systems. Section~\ref{methodology} presents the proposed hierarchical GNN framework and graph construction methodology. Section~\ref{sec:qos_alg} details the QoS awareness assignment algorithm and QoS-aware training strategies. Section~\ref{sec:evaluation} provides comprehensive experimental evaluation and results analysis. The paper concludes in Section~\ref{sec:conclusion} with future research directions.

\section{Related Work}\label{sec:related_work}

\subsection{AI-based Network Traffic Classification} 

Prior research in NTC has laid a strong theoretical foundation from multiple perspectives. Table~\ref{table:NTC_comparison} summarizes representative research on NTC, focusing on recent AI techniques. As it shows, recent AI-based solutions have demonstrated near-perfect accuracy (typically around 90\%) in basic app identification even on encrypted traffic~\cite{9882011,zhao2023yet,9319399,9933044}. Among these approaches, GraphDapp~\cite{9319399} models traffic flows as graph structures, where nodes represent network endpoints and edges capture communication patterns, enabling effective app identification with 89\% accuracy through graph neural networks. ProGraph~\cite{9933044} extends graph-based approaches by incorporating protocol-level features and achieving over 92\% accuracy in distinguishing between different applications under distinct networking scenarios. Parallel efforts in network intrusion and anomaly detection have also achieved high accuracy rates ($>$95\%)~\cite{zhao2024towards, 10190520, 10179289, wang2023bars, 9488690, yang2021cade}. CADE~\cite{yang2021cade} employs contrastive learning to detect adversarial attacks in encrypted traffic, achieving over 95\% detection accuracy by learning robust feature representations. ACID~\cite{9488690} improves model robustness against evasion attacks, demonstrating 99\% accuracy in identifying malicious traffic patterns. BARS~\cite{wang2023bars} specifically addresses the robustness of NTC systems against adversarial perturbations. However, most existing work focuses on coarse-grained app-level labeling. In practice, traffic patterns from the same app can be highly heterogeneous, reflecting the contextual complexity of service behaviors. As a result, while these solutions provide a solid foundation, they often fall short in supporting QoS provisioning or resource management in edge network environments.

\begin{table}[ht]
    \footnotesize
      \renewcommand{\arraystretch}{.9}
    \centering        
    \vspace{+2mm}
    \caption{A comparison summary of selected prior literature.}\label{table:NTC_comparison}
    \begin{tabular}{c|c|c|c|c}
    \toprule
      \textbf{Algorithm}  & \textbf{Flow} & \textbf{Target} & \textbf{Acc.}  & \textbf{QoS} \\      
      \midrule
       AI-NTC~\cite{9882011} & NA & app label & $>$90\% &  No \\ 
       
       ET-BERT~\cite{zhao2023yet} & $\uparrow\downarrow$ & app label & $>$92\%  & No \\
       GraphDapp~\cite{9319399} & $\uparrow\downarrow$ & app label & $>$89\%  & No \\
       ProGraph~\cite{9933044} & $\uparrow\downarrow$ & app label& $>$90\%  & No\\
        \hline
        CADE~\cite{yang2021cade} & $\uparrow\downarrow$ & Attacks & $>$95\%  & NA \\
        ACID~\cite{9488690} &  $\uparrow\downarrow$ & Attacks & $>$99\%  & NA\\  
        AWEE~\cite{10870109} & NA & Attacks & $>$98\%  & NA\\        
         BARS~\cite{wang2023bars} & NA & Robustness & NA  & NA \\
       \midrule
       P2P-act~\cite{5936162} & $\downarrow$ & P2P actions & NA & No\\
        Web-act~\cite{7020869} & NA & Web actions & $>$92\% & No  \\
        CUMMA~\cite{7377112} & $\downarrow$ & MSG services & $>$90\%  & Yes \\    
        rCKC+FRF~\cite{10.1145/3097983.3098049} & $\uparrow\downarrow$ & MSG/SM services & $>94\%$ & Yes \\
       \bottomrule       
    \end{tabular}    
    \label{tab:channel_models}
    \vspace{-4mm}
\end{table}
\subsection{Service Aware Network Traffic Classification} 

To bridge the gap between coarse-grained application classification and service-level differentiation, researchers have explored fine-grained NTC methods over the past decade. Early efforts focused on identifying functional categories within specific applications. For example, Park et al.~\cite{5936162} proposed a method to classify peer-to-peer (P2P) traffic by activity type (e.g., download, upload, and search) using Jaccard similarity. Their analysis showed that downloading traffic dominated usage on platforms such as Fileguri and BitTorrent, accounting for 74\%–90\% of total traffic. Lin et al.~\cite{7020869} extended this approach to web applications. They classified user actions such as video streaming and map browsing by analyzing statistical features from HTTPS messages without relying on payload inspection. Their method achieved up to 98.30\% accuracy. Fu et al.~\cite{7377112} focused on mobile messaging applications such as WeChat and WhatsApp. By combining packet- and flow-level features, they classified activities like text messaging and voice calls with over 90\% accuracy. Liu et al.~\cite{10.1145/3097983.3098049} developed a real-time analysis framework for encrypted mobile traffic. Their method achieved 94.01\% accuracy on WeChat while significantly improving processing speed and memory efficiency.
Despite these advances, scaling fine-grained classification across a broad range of applications in dynamic, heterogeneous edge environments remains an open challenge.

\section{Framework of the Hierarchical Graph Neural Network based NTC}\label{methodology}

\begin{figure}[b]
    \centering
    \includegraphics[width=.9\linewidth]{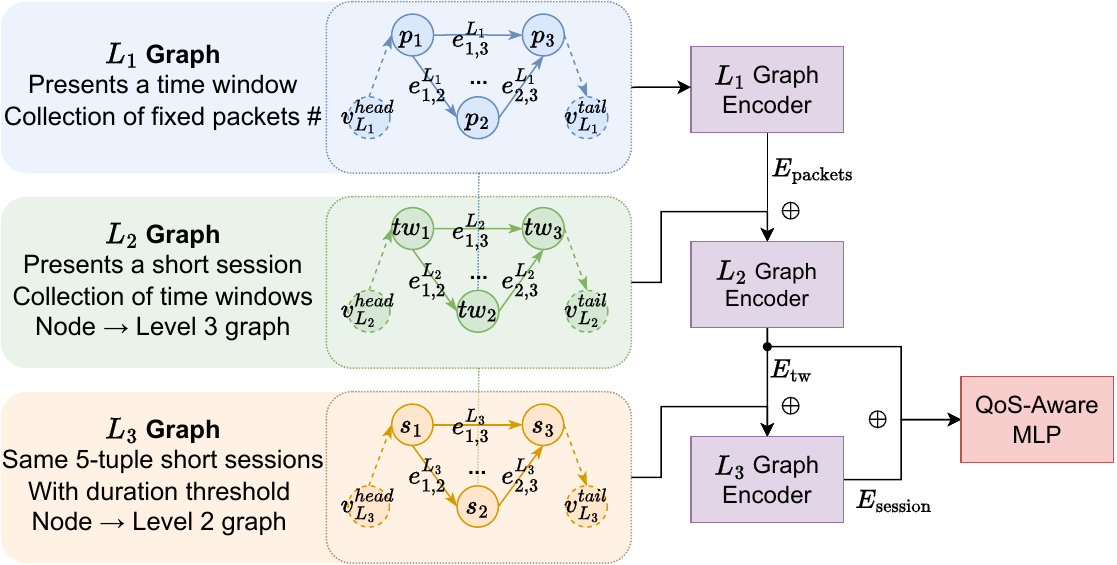}
    \caption{Overview of the hierarchical GNN framework. }
    \label{fig:GNN_overview}
\end{figure}

The overall architecture of the proposed hybrid Graph Neural Network (GNN) model is depicted in Fig.~\ref{fig:GNN_overview}. The model is designed to capture multi-scale structural and temporal dependencies inherent in network traffic through a three-tiered hierarchical encoding framework, followed by a unified classification module.

\subsection{Graph Construction}\label{data process}

Network packets are initially grouped based on the canonical 5-tuple: \textit{source IP address, source port, destination IP address, destination port, and protocol}. The source and destination IP addresses are considered interchangeable for the bidirectional flows in the same session. A session timeout threshold (e.g., 0.5 seconds) is applied to segment prolonged flows into shorter sessions, while a maximum session duration (e.g., 60 seconds) is enforced to bound session length. To model the hierarchical and temporal structure of network traffic, we construct a three-level graph representation that captures traffic characteristics at multiple granularities: \textit{packet aggregation}, \textit{time windowing}, and \textit{session clustering}. As illustrated in Fig.~\ref{fig:GNN_overview}, the graph construction proceeds in three stages, each corresponding to a distinct level of abstraction. The architecture incorporates 18 semantic features (Table~\ref{tab:feature_definitions}) that encode statistical and temporal properties across these levels, enabling the model to learn expressive, multi-scale representations for QoS-aware traffic classification.

\textit{Level-1 Nodes (Packet Aggregation):} 
Within each Level-2 time window, packets are grouped into Level-1 nodes based on a fixed packet count (e.g., 10 packets per node). Each Level-1 node is represented by a 9-dimensional feature vector capturing fine-grained traffic characteristics, including basic statistical metrics and inter-arrival timing patterns (Table~\ref{tab:feature_definitions}). Higher-order distributional features are excluded at this level. Each packet cluster forms an independent Level-1 subgraph.

\begin{table}[ht]
\centering
\footnotesize
\vspace{+2mm}
\caption{Multi-level feature definitions. }
\label{tab:feature_definitions}
\renewcommand{\arraystretch}{1.3} 
\begin{tabular}{|p{0.17\linewidth}|p{0.47\linewidth}|p{0.03\linewidth}|p{0.03\linewidth}|p{0.03\linewidth}|}
\hline
{\centering\textbf{Feature}} & {\centering\textbf{Notation and calculation}} & \textbf{L1} & \textbf{L2} & \textbf{L3} \\
\hline
\# packet & $n$ & $\checkmark$ & $\checkmark$ & $\checkmark$ \\
\hline
Total bytes & $\sum_{i=1}^{n} l_i$ & $\checkmark$ & $\checkmark$ & $\checkmark$ \\
\hline
Mean(bytes) & $\frac{\sum_{i=1}^{n} l_i}{n} = \overline{l}$ & $\checkmark$ & $\checkmark$ & $\checkmark$ \\
\hline
Var(bytes) & $\frac{1}{n}\sum_{i=1}^{n}(l_i - \overline{l})^2 = \sigma_l^2$ & $\checkmark$ & $\checkmark$ & \\
\hline
Uplink ratio & $\frac{1}{n}\sum_{i=1}^{n}\mathbf{1}(\text{src}_i = \text{client\_ip})$ & $\checkmark$ & $\checkmark$ & $\checkmark$ \\
\hline
Mean(IAT) & $\frac{1}{n-1}\sum_{k=2}^{n}(t_k - t_{k-1}) = \overline{\mathrm{IAT}}$ & $\checkmark$ & $\checkmark$ & \\
\hline
Var(IAT) & $\frac{1}{n-1}\sum_{k=2}^{n}(\mathrm{IAT}_k - \overline{\mathrm{IAT}})^2$ & $\checkmark$ & $\checkmark$ & \\
\hline
Min(IAT) & $\min_{2\le k\le n}\mathrm{IAT}_k$ & $\checkmark$ & $\checkmark$ & \\
\hline
Max(IAT) & $\max_{2\le k\le n}\mathrm{IAT}_k$ & $\checkmark$ & $\checkmark$ & \\
\hline
Skewness & $\begin{cases}
\displaystyle\frac{\frac{1}{n}\sum_{i=1}^{n}(l_i - \overline{l})^3}{(\sigma_l^2)^{3/2}}, & \sigma_l^2 > 0\\
0, & \text{else}
\end{cases}$ & & $\checkmark$ & \\
\hline
Kurtosis & $\begin{cases}
\displaystyle\frac{\frac{1}{n}\sum_{i=1}^{n}(l_i - \overline{l})^4}{(\sigma_l^2)^2} - 3, & \sigma_l^2 > 0\\
0, & \text{else}
\end{cases}$ & & $\checkmark$ & \\
\hline
Session dur. & $t_{\mathrm{end}} - t_{\mathrm{start}}$ & & & $\checkmark$ \\
\hline
Packet rate & ${{n}}/({t_{\mathrm{end}} - t_{\mathrm{start}}})$ & & & $\checkmark$ \\
\hline
Byte rate & $({\sum_{i=1}^{n} l_i})/({t_{\mathrm{end}} - t_{\mathrm{start}}})$ & & & $\checkmark$ \\
\hline
Flow symm. & $1 - \frac{|\overline{l}_{\text{up}} - \overline{l}_{\text{down}}|}{\max(\overline{l}_{up},\,\overline{l}_{down})}$ & & & $\checkmark$ \\
\hline
Burst count & $|\{B_i : \mathrm{IAT}\le 100\text{ms}\}|$ & & & $\checkmark$ \\
\hline
Mean(burst) & $\frac{1}{B}\sum_{i=1}^{B}|\text{burst}_i|$ & & & $\checkmark$ \\
\hline
Burst interval  & $\frac{1}{B-1}\sum_{i=1}^{B-1}(t_{\mathrm{start},i+1}-t_{\mathrm{end},i})$ & & & $\checkmark$ \\
\hline
\textbf{\# Features} & & \textbf{9} & \textbf{11} & \textbf{11} \\
\hline
\end{tabular}%
\end{table}
\renewcommand{\arraystretch}{1.0}

\textit{Level-2 Nodes (Time Window Clusters):} 
Within each short session (segmented using a fixed idle timeout, e.g., 0.5 second), non-empty time windows (e.g., 100 ms) are aggregated into Level-2 nodes. Each node represents a time window cluster and forms an independent Level-1 subgraph. Each Level-2 node is encoded with an additional 11-dimensional feature vector comprising nine shared features and two higher-order statistical features—skewness and kurtosis—computed as described in Table~\ref{tab:feature_definitions}. These features provide medium-grained temporal insights and capture the distributional characteristics of packet lengths within each time window.

\textit{Level-3 Nodes (Session Aggregation):} 
Multiple short sessions associated with the same 5-tuple are aggregated into Level-3 nodes. To ensure compatibility with real-time constraints, each Level-3 session is limited to a fixed maximum duration (e.g., 60 seconds), with longer sessions split accordingly. Beyond the embedding features from the Level-2 subgraph, each Level-3 node is encoded with an addition 11-dimensional feature vector, including four shared features (packet count, total bytes, average packet size, uplink ratio) and seven session-specific features (session duration, packet rate, byte rate, flow symmetry, burst count, average burst size, inter-burst time), as detailed in Table~\ref{tab:feature_definitions}. These features abstract long-term behavioral patterns and emphasize burst-level dynamics and flow characteristics.

To address the challenge of graphs containing only a single real node, where GNNs struggle due to the absence of neighborhood context, auxiliary head and tail nodes are introduced at all levels. These auxiliary nodes are assigned zero-valued feature vectors with dimensionality matching that of the corresponding real nodes (9-dimensional for Level-1, 11-dimensional for Level-2 and Level-3, respectively).

\textit{Intra-level Edges:} 
Within each level, nodes are fully connected in forward temporal order, with edge weights reflecting time delays between consecutive nodes $i$ and $j$:
\begin{subequations}\label{eq:edge}
\begin{align}
\text{edge\_weight}^{L_1}_{i,j} &= \text{timestamp}_j - \text{timestamp}_i, \\
\text{edge\_weight}^{L_2}_{i,j} &= \text{center\_time}_j - \text{center\_time}_i, \\
\text{edge\_weight}^{L_3}_{i,j} &= \text{session\_start}_j - \text{session\_end}_i.
\end{align}
\end{subequations}
These edge weights encode temporal dependencies: Level-1 edges capture delays between packet aggregations, Level-2 edges capture delays between time window centers, and Level-3 edges capture inter-session gaps. Zero-weight edges connect auxiliary head and tail nodes to the first and last real nodes, respectively, ensuring structural consistency.

\textit{Inter-level Edges:} 
The hierarchical structure maintains strict correspondence across levels without explicit inter-level edges. Each Level-3 session node aggregates multiple Level-2 time window subgraphs, and each Level-2 node aggregates multiple Level-1 packet cluster subgraphs. Information propagates bottom-up through learned feature embeddings: Level-1 features inform Level-2 representations, which in turn inform Level-3 behavioral abstractions.

This hierarchical design enables the model to capture multi-scale temporal patterns-ranging from fine-grained packet-level interactions (Level-1), through medium-grained time window dependencies (Level-2), to coarse-grained session-level behaviors (Level-3), while preserving temporal ordering and causal relationships at each level.

\subsection{Hierarchical Graph Encoder and QoS-aware Classifier}

\begin{figure}[ht!]
    \centering
    \includegraphics[width=1\linewidth]{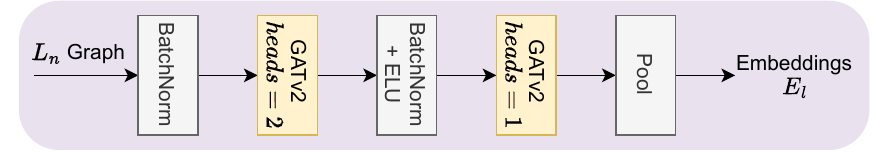}
    \caption{Overview of the graph encoder.}
    \label{fig:graph_encoder}
\end{figure}

A sub-graph in each level is processed by a 2-layer graph encoder based on based on GATv2~\cite{DBLP:journals/corr/abs-2105-14491}, as depicted in Fig.~\ref{fig:graph_encoder}. The designs of the graph encoder are slightly different, described in the following.
\begin{itemize}
    \item The level-1 graph encoder employs 2 attention heads with edge feature integration in the first layer, transforming input features to 64-dimensional representations. The second layer uses single-head attention to produce 64-dimensional node embeddings. Dual global pooling operations (mean and max) aggregate node representations into 128-dimensional cluster embeddings.
    
    \item The level-2 graph encoder processes the 11-dimensional features from time window nodes, and the 128-dimensional Level-1 cluster embeddings, creating 139-dimensional features. The augmented features undergo 2-layer GATv2 processing: the first layer with 2 attention heads expands to 256 dimensions, while the second layer with single attention consolidates to 128-dimensional embeddings. Global pooling produces 256-dimensional time window representations.

    \item The level-3 graph encoder process the 11-dimensional features from session nodes, and the 256-dimensional Level-2 embeddings, creating 267-dimensional features. Similar 2-layer GATv2 processing expands features to 256 dimensions, then consolidates to 128-dimensional session embeddings. Global pooling yields 256-dimensional session-level representations.

\end{itemize}

\begin{figure}[ht!]
    \centering
    \includegraphics[width=1\linewidth]{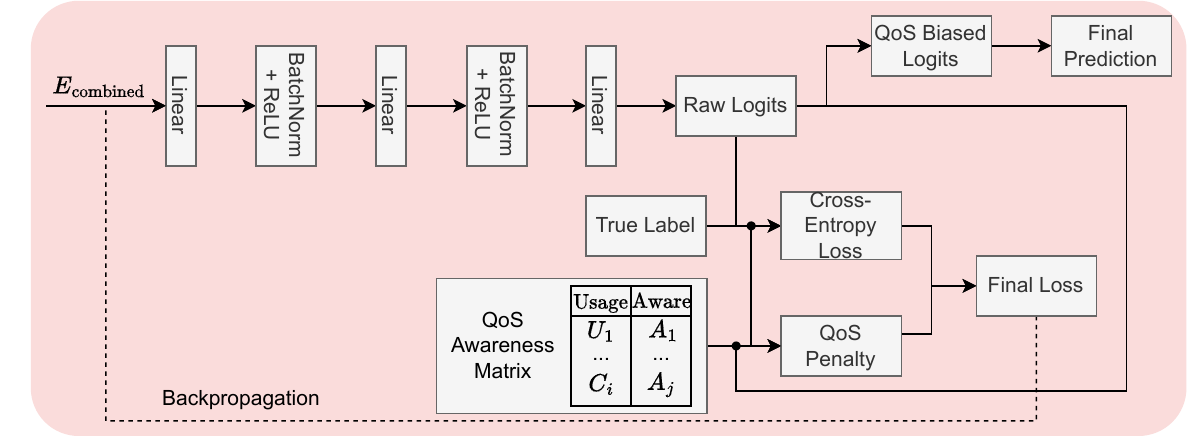}
    \caption{Overview of the QoS-aware classifier. }
    \label{fig:qos_classifer}
\end{figure}

The final stage of the model is a QoS-aware classification network designed for fine-grained, traffic categorization at the Level-2. As shown in Fig.~\ref{fig:qos_classifer}, the classifier leverages a multi-scale feature fusion strategy, combining context from both the Time Window (TW) and its parent Session to make a prediction. For each TW graph to be classified, its learned 256-dimensional embedding, $\mathbf{E}_{\text{tw}}$, is concatenated with the 256-dimensional embedding of its corresponding parent Session, $\mathbf{E}_{\text{session}}$, which contains the information learning from all three level. This creates a combined 512-dimensional feature vector, $\mathbf{E}_{\text{combined}} = [\mathbf{E}_{\text{tw}} \mathbin\Vert \mathbf{E}_{\text{session}}]$, that encapsulates both temporal patterns directly from the TW and behavioral context from the all three levels. This combined embedding is then passed through a Multi-Layer Perceptron (MLP) which acts as the classifier:
\begin{enumerate}
    \item A linear layer maps the 512-dimensional input to a 512-dimensional hidden space, followed by Batch Normalization, a ReLU activation, and Dropout.
    \item A second linear layer reduces the dimensionality from 512 to 256, again followed by Batch Normalization, ReLU, and Dropout.
    \item A final linear output layer maps the 256-dimensional representation to a $C$-dimensional logit vector, where $C$ is the number of the classes.
\end{enumerate}
The resulting logits are used to compute the classification loss and final predictions.

\section{QoS Awareness NTC}\label{sec:qos_alg}

\subsection{QoS Awareness Assignment}

To align the model's predictions with network Quality of Service (QoS) requirements, we introduce a QoS-aware training and inference framework. This framework prioritizes the correct classification of high QoS awareness traffic (e.g., live streaming) and penalizes misclassifications that would lead to assigning a lower-than-required service level.
QoS awareness assignment employs a magnitude-based approach that automatically determines awareness levels for network traffic flows based on their service requirements. The algorithm analyzes traffic characteristics using logarithmic magnitude classification and weighted scoring to establish differentiated service awareness.
For better illustration, the QoS awareness of each traffic flow $f$ is characterized by five QoS metrics $c_i$: \emph{bandwidth \text{(Mbps)}}, \emph{jitter stability}, \emph{packet stability}, \emph{average inter-burst delay}, and \emph{burst stability}, denoted as $[c_{\text{bw}}, c_{\text{jitter}}, c_{\text{packet}}, c_{\text{burst\_freq}}, c_{\text{burst\_stab}}]$, where 
\begin{subequations}\label{eq:QoSMetrics}
\begin{align}
   & c_{\text{bw}}   = \frac{\sum_{i=1}^{N} \text{size}_i \times 8}{(t_{\text{end}} - t_{\text{start}}) \times 10^6} , \\
   &  c_{\text{jitter}}        = \frac{\sigma_{\text{IAT}}}{\mu_{\text{IAT}}}, \\
   & c_{\text{packet}}        = \sigma_{\text{IAT}}, \\
  & c_{\text{burst\_freq}}  = \frac{1}{N_{\text{bursts}}-1} \sum_{k=1}^{N_{\text{bursts}}-1} (t_{\text{start}}^{k+1} - t_{\text{end}}^{k}), \\
   & c_{\text{burst\_stab}}          = \sigma_{\text{inter\_burst\_delay}}.
\end{align}
\end{subequations}
For each QoS metric $c_i$ ($i$ being an index to the QoS metric, e.g., $c_1$ indicates $c_\text{bw}$), we perform a logarithmic transformation to normalize the scale and emphasize order-of-magnitude differences:
\begin{equation}
m_i = \log_{10}(c_i), \quad i~\text{is indexed to QoS metrics}.
\end{equation}
The logarithmic transformation is applied directly, as it preserves order-of-magnitude distinctions across the full range of metric values, with $m_i$ being negative for fractional values and positive for values greater than 1.
The classification process operates independently for each QoS metric, grouping traffic flows based on magnitude similarity. For each metric, all flows are first sorted by their transformed magnitude values $m_i$ in ascending order. The classification then proceeds sequentially through this sorted list:
\begin{itemize}
\item \textbf{Class Initiation:} The first flow in the sorted list initializes the first QoS class (e.g., Class~\textbf{0}) for the current metric. Its transformed magnitude $m_i^{(1)}$ serves as the reference point for subsequent comparisons.
\item \textbf{Sequential Assignment:} For each subsequent flow $k$ in the sorted order, its magnitude $m_i^{(k)}$ is compared against the magnitude of the most recently processed flow in the current class. If the magnitude difference satisfies:
\begin{equation}
|m_i^{(k)} - \text{median}(\{m_i^{(j)}: j \in \text{Class}\})| \leq X_{\text{thresh}},
\end{equation}
where $\text{median}(\{m_i^{(j)}: j \in \text{Class}\})$ is the median magnitude of all flows currently in that class, then flow $k$ is assigned to that class.

\item \textbf{New Class Creation:} If the magnitude difference exceeds the threshold (i.e., $|m_i^{(k)} - \text{median}(\{m_i^{(j)} : j \in \text{Class}\})| > X_{\text{thresh}}$), a new QoS class is created, and flow $k$ becomes the first member of this new class.

\end{itemize}
This process is repeated independently for all five QoS metrics: bandwidth, jitter, packet stability, burst frequency, and burst stability. Each metric produces its own set of classes, and each traffic flow receives a class assignment for every metric.

\noindent\textit{Illustrative example: } Consider a bandwidth metric with three flows having transformed magnitudes $m_{\text{bw}}^{(1)}$, $m_{\text{bw}}^{(2)}$, and $m_{\text{bw}}^{(3)}$ where $m_{\text{bw}}^{(1)} < m_{\text{bw}}^{(2)} < m_{\text{bw}}^{(3)}$. In this work, we set $X_{\text{thresh}} = 0.6$ to capture one order-of-magnitude differences between classes. The example followed the setting. Flow $f_1$ initiates Class~\textbf{0}. Flow $f_2$ is compared: if $|m_{\text{bw}}^{(2)} - m_{\text{bw}}^{(1)}| \leq 0.6$, it joins Class~\textbf{0}. Flow $f_3$ is compared against the last assigned flow: if $|m_{\text{bw}}^{(3)} - median(m_{\text{bw}}^{(1)},m_{\text{bw}}^{(2)})| > 0.6$, it creates a new Class~\textbf{1}.

After independent classification of each metric, every traffic flow is characterized by a 5-dimensional class sequence vector $\mathbf{s} = [s_1, s_2, s_3, s_4, s_5]$, where $s_i$ represents the class assignment for the $i$-th QoS metric (bandwidth, jitter, packet stability, burst frequency, burst stability, respectively). Traffic flows with identical class sequence vectors are grouped into the same QoS label. After initial classification, QoS classes are reordered based on their relative importance in network management. For instance, higher bandwidth and lower jitter typically indicate higher QoS priority. To formalize this, we define a QoS awareness score $p_\text{a}^{\textbf{k}}$ for each class $\textbf{k}$ based on its class sequence values:
\begin{equation}
p_\text{a}^{\textbf{k}} = w_1 \cdot {s}_1^{\textbf{k}} + \sum_{i=2}^{5} w_i \cdot \left( (\max_{j \in \textbf{k}} s_i^{(j)}) - {s}_i^{\textbf{k}} \right),
\end{equation}
where the first term rewards higher bandwidth (QoS metric $s_1$), and the remaining terms penalize instability in jitter, packet size, inter-burst delay, and burstiness metrics, where lower values indicate better QoS. The weights $w_i$ can be tuned based on application-specific requirements. In this work, we prioritize real-time responsiveness and assign the weights as follows: $w_{\text{bandwidth}} = 0.30$, $w_{\text{jitter}} = 0.20$, $w_{\text{packet}} = 0.15$, $w_{\text{burst\_freq}} = 0.20$, and $w_{\text{burst\_stab}} = 0.15$. 
Classes are then ranked in ascending order of $p_\text{a}^{\textbf{k}}$, with higher scores indicating higher QoS awareness. The final QoS levels are assigned accordingly, ranging from $0$ to $N-1$ for $N$ classes.
This magnitude-based classification framework automatically adapts to diverse traffic distributions; ensures similar flows are grouped under the same QoS class; and provides interpretable and tunable prioritization based on weighted QoS metrics.

\subsection{QoS-aware Model Training and Inference}

To incorporate QoS awareness into the training process, we design a composite loss function that balances standard classification accuracy with penalties for QoS-violating misclassifications. The total loss is defined as:
\begin{equation}
\mathcal{L}_{\text{total}} = (1-\lambda) \mathcal{L}_{\text{CE}} + \lambda \mathcal{L}_{\text{QoS}},
\end{equation}
where $\mathcal{L}_{\text{CE}}$ is the standard cross-entropy loss, $\mathcal{L}_{\text{QoS}}$ is the QoS-aware penalty term, and $\lambda \in [0,1]$ is a tunable hyperparameter that controls the trade-off between classification accuracy and QoS sensitivity.
The QoS-aware loss $\mathcal{L}_{\text{QoS}}$ is computed by scaling the cross-entropy loss with a penalty matrix $P[i,j]$ that encodes the cost of misclassifying a sample from class $i$ as class $j$:
\begin{equation}
\mathcal{L}_{\text{QoS}} = \mathcal{L}_{\text{CE}} \times (1 + P[y_{\text{true}}, y_{\text{pred}}]),
\end{equation}
where the penalty matrix $P$ is defined as follows:
\begin{itemize}
    \item $P[i,i] = 0$ for correct classifications.
    \item $P[i,j] = \beta$ for misclassifications to higher or equal QoS classes, i.e., $\text{QoS}(j) \geq \text{QoS}(i)$.
    \item $P[i,j] = 1.0 + \gamma \cdot (\text{QoS}(i) - \text{QoS}(j))$ for misclassifications to lower QoS classes.
\end{itemize}
Here, $\beta$ and $\gamma$ are hyperparameters that control the severity of penalties, with $\gamma$ typically set higher to discourage under-provisioning errors.

To further align predictions with QoS priorities during inference, we introduce three complementary strategies. 

\textit{QoS bias adjustment: }The raw output logits are adjusted by incorporating a bias term proportional to each class's QoS awareness score:
\begin{equation}
\text{logits}_{\text{biased}} = \text{logits}_{\text{raw}} + \alpha \cdot \text{QoS}_{\text{awareness}},
\end{equation}
where $\alpha$ is a tunable parameter that controls the strength of the QoS bias. This encourages the model to favor higher-QoS classes when confidence is comparable.

\textit{Post-processing refinement: }For predictions with low confidence (i.e., maximum softmax probability $\text{score}_{\text{top-1}}$ below a threshold $\sigma$), we compare the top-2 candidate classes. If their confidence scores are within a relative margin $\theta$, the class with the higher QoS awareness score is selected:
\begin{equation}
\frac{\text{score}_{\text{top-2}}}{\text{score}_{\text{top-1}}} < \theta \quad \Rightarrow \quad \text{select class with higher QoS}.
\end{equation}

\textit{QoS-aware evaluation metrics: }In addition to standard accuracy, we introduce two QoS-centric evaluation metrics:
\begin{itemize}
    \item \text{QoS satisfaction rate:} The percentage of samples where the predicted QoS level is greater than or equal to the ground truth in misclassified samples.
    \item \text{QoS experience score:} A metric that rewards over-provisioning errors (predicting higher QoS than required) more than under-provisioning errors.
\end{itemize}

This QoS-aware training and inference framework ensures that classification errors are biased toward over-provisioning rather than under-provisioning, thereby preserving service quality for latency-sensitive or mission-critical applications. It provides a principled mechanism to integrate application-level QoS priorities into both model optimization and decision-making, ultimately enhancing the reliability and utility of traffic classification in real-world network environments.

\section{Evaluation Results}\label{sec:evaluation}

A comprehensive evaluation is conducted on the proposed QoS-aware hierarchical GNN model for fine-grained network traffic classification. The experiments demonstrate the effectiveness of the three-level graph representation and the QoS-integrated training strategy in accurately classifying 14 traffic classes across four widely used applications.

\subsection{Data Collection}

The dataset used in this study was collected using \texttt{PCAPdroid}~\cite{pcapdroid} on Android devices connected to WiFi networks, capturing real-world traffic traces from four major applications: \textit{YouTube}, \textit{Prime Video}, \textit{TikTok}, and \textit{Zoom}. For each application, 10-minute PCAPNG traces were recorded under diverse usage scenarios to construct a comprehensive 14-class dataset:
\begin{itemize}
    \item \textbf{YouTube}: Browsing, live streaming, long-form video, short-form video (4 classes).
    \item \textbf{Prime Video}: browsing, live streaming, long-form video (3 classes).
    \item \textbf{TikTok}: Browsing, live streaming, short-form video (3 classes).
    \item \textbf{Zoom}: Audio conferencing, symmetric video conferencing, uplink-only presentation mode, downlink-only attendance mode (4 classes).
\end{itemize}
Raw packet traces were processed to extract sessions using 5-tuple flow identification and idle timeout segmentation. Each session was then transformed into a three-level hierarchical graph structure comprising:
\begin{itemize}
    \item \textbf{Level-1 (packet cluster graphs)}: Nodes represent packet clusters aggregated by fixed packet count.
    \item \textbf{Level-2 (time window graphs)}: nodes represent 100\,ms time windows within short sessions.
    \item \textbf{Level-3 (session graphs)}: Nodes represent short sessions grouped under the same 5-tuple, constrained to a maximum duration of 60 seconds.
\end{itemize}
The resulting dataset exhibits natural class imbalance, reflecting realistic usage distributions and providing a challenging yet authentic benchmark for evaluating classification performance in practical network environments.

\begin{figure*}[ht!]
    \centering
    \subfigure[The 3-level hierarchical graph.]{\includegraphics[width = 0.7\linewidth]{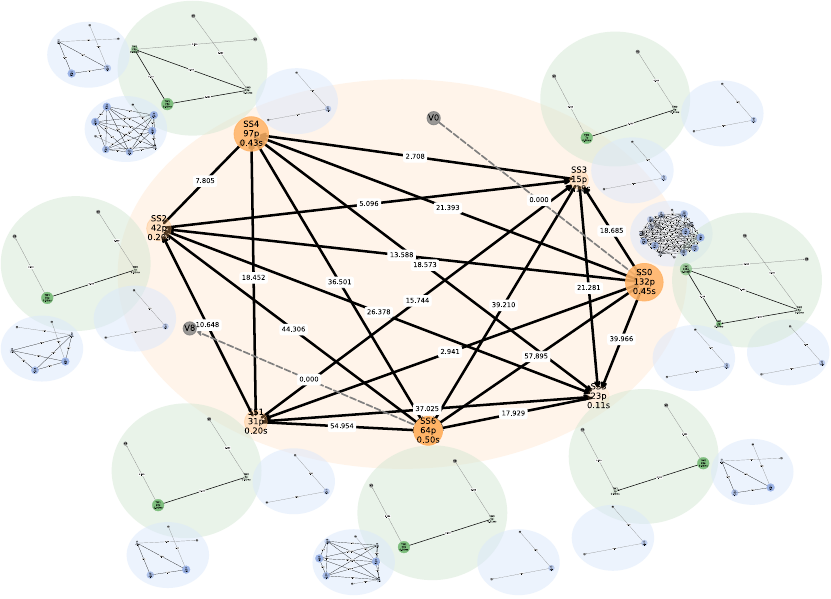}}\\
    \subfigure[The I/O graph of the session.]{\includegraphics[width = 0.7\linewidth]{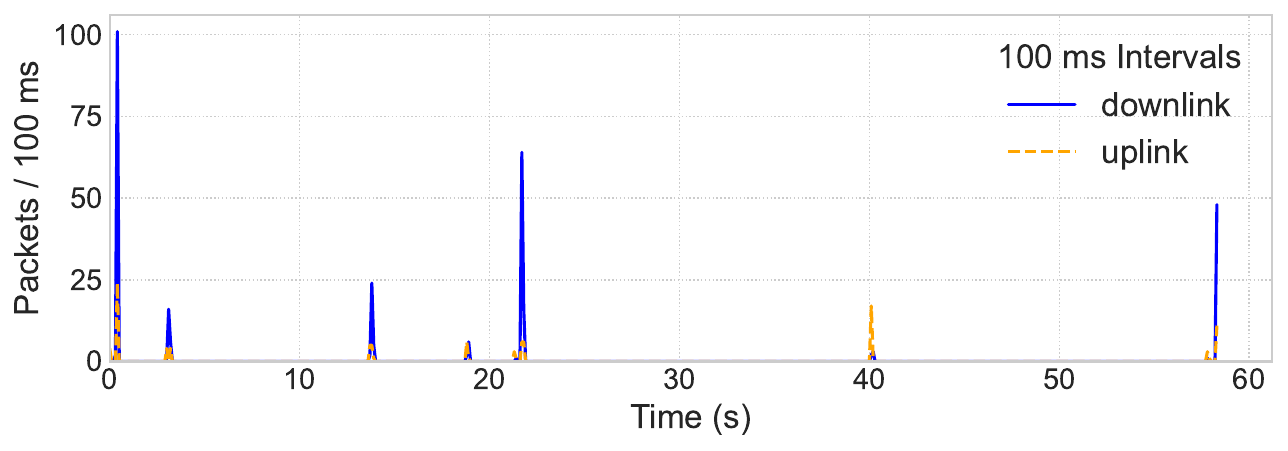}}
\caption{Example of the multi-level graph structure and the corresponding raw session traffic.} \label{fig:graph_example}    \end{figure*}

Fig.~\ref{fig:graph_example} illustrates an example of a 3-level hierarchical graph from YouTube Browsing traffic. As shown in Fig.~\ref{fig:graph_example}(a), blue nodes represent Level-1 packet cluster graphs, green nodes represent Level-2 time window graphs, and 
orange nodes represent Level-3 session graphs. Gray nodes indicate auxiliary nodes; solid arrows denote real temporal edges with time-delay labels; dashed arrows connect virtual nodes. Node size encodes total bytes, and transparency reflects session duration (Level-3) or average packet length (Level-1 and Level-2).
Fig.~\ref{fig:graph_example}(b) shows the I/O traffic graph of the original session corresponding to Fig.~\ref{fig:graph_example}(a), showing the temporal network activity used to construct the hierarchical graph.

\subsection{Experimental Setup and Evaluation Metrics}

The experimental evaluation follows an 80/20 train-test split using stratified sampling to preserve the original class distribution across both subsets. The model is implemented using the PyTorch Geometric framework and optimized with the AdamW optimizer. A learning rate scheduler is employed to ensure stable convergence during training. We first evaluate a conventional Packet-level Multi-Layer Perceptron NTC which has been used in~\cite{9882011}, to do in-app traffic classification, which achieves only 72.7\% accuracy, indicating that conventional methods cannot effectively distinguish in-app traffic. Therefore, we employ our proposed QoS-aware hierarchical GNN approach to address these limitations. To isolate the impact of QoS-awareness, two models are trained and evaluated under identical conditions: (1) A baseline model without QoS-aware loss or inference strategies; and (2) The proposed QoS-aware hierarchical GNN model. Both models utilize the same dataset, preprocessing pipeline, and data splits, ensuring a controlled comparison. 
The experimental evaluation follows an 80/20 train-test split using stratified sampling to preserve the original class distribution across both subsets. The model is implemented using the PyTorch Geometric framework and optimized with the AdamW optimizer. A learning rate scheduler is employed to ensure stable convergence during training. Therefore, we employ our proposed QoS-aware hierarchical GNN approach to address these limitations. To isolate the impact of QoS-awareness, two models are trained and evaluated under identical conditions: (1) A baseline model without QoS-aware loss or inference strategies; and (2) The proposed QoS-aware hierarchical GNN model. Both models utilize the same dataset, preprocessing pipeline, and data splits, ensuring a controlled comparison.

Model performance is assessed using both conventional and QoS-centric evaluation metrics. The conventional metrics focus on traditional classification accuracy, measuring the overall correctness of predicted traffic classes. The accuracy performance is evaluated using standard classification metrics: Precision, Recall, and F1-Score. The first QoS evaluation metric is \textit{QoS satisfaction rate}, which quantifies the proportion of predictions where the predicted QoS level is greater than or equal to the ground truth in the misclassified samples, reflecting over-provisioning behavior.   
To further evaluate the effectiveness of QoS-aware classification, a new metric \textit{QoS experience score} is introduced. This metric extends beyond traditional accuracy by incorporating the severity of misclassifications based on QoS awareness levels, thereby assessing the practical impact of prediction errors in network traffic management.
The QoS experience score is computed using a reward-penalty mechanism applied to the confusion matrix:
\begin{equation}
\mathcal{Q}_{\text{score}} = \sum_{i=1}^{N} \sum_{j=1}^{N} C_{i,j} \cdot w_{i,j},
\end{equation}
where $C_{i,j}$ denotes the number of samples with true class $i$ predicted as class $j$, and $w_{i,j}$ is the weight assigned to each prediction outcome:
\begin{equation}
w_{i,j} = \begin{cases}
+P_i, & \text{if } P_j \geq P_i, \quad \text{over-provisioning bias}, \\
-P_i, & \text{if } P_j < P_i, \quad \text{under-provisioning bias},
\end{cases}
\end{equation}
where $P_i$ denotes the QoS awareness level of class $i$. The scoring logic is as follows:
\begin{itemize}
    \item \textbf{Over-provisioning bias} ($P_j \geq P_i$): When a flow is classified into a class with equal or higher QoS awareness than its true class, is it more likely to allocate sufficient resources with over provisioning, earning a positive score proportional to the true class's awareness level.
    
    \item \textbf{Under provisioning bias} ($P_j < P_i$): When high-awareness traffic is misclassified into a lower-awareness class, it risks resource under-provisioning that cannot meet QoS needs, incurring a penalty proportional to the true class's awareness level.
\end{itemize}
The theoretical maximum score, representing perfect classification, is given by:
\begin{equation}
\mathcal{Q}_{\text{max}} = \sum_{i=1}^{N} n_i \cdot P_i,
\end{equation}
where $n_i$ is the number of samples in class $i$. The QoS Score Ratio provides a normalized performance metric:
\begin{equation}
\mathcal{Q}_{\text{ratio}} = \frac{\mathcal{Q}_{\text{score}}}{\mathcal{Q}_{\max}} \times 100\%.
\end{equation}
This ratio provides a meaningful comparison between QoS-aware and conventional models, reflecting the practical consequences of misclassification in network resource allocation. Higher QoS scores indicate better alignment with service-level requirements, while lower scores suggest potential degradation due to inappropriate traffic prioritization.

\begin{table*}[ht]
\centering
\caption{QoS metrics for adaptive awareness assignment.}
\label{tab:qos_metrics}
\footnotesize
\begin{tabular}{@{}lccccccc@{}}
\toprule
\textbf{Application} & \textbf{Bandwidth} & \textbf{Jitter Stability} & \textbf{Packet Stability} & \textbf{Burst Frequency} & \textbf{Burst Stability} & \textbf{Class} & \textbf{QoS} \\
\textbf{Usage Type}
 & \textbf{ (Mbps)} & \textbf{(CV)} & \textbf{(ms)} & \textbf{(ms)} & \textbf{(ms)} &  \textbf{Sequence} &  \textbf{Awareness} \\
\midrule
Prime Video Browse & 1.526 & 16.878 & 264.646 & 819.475 & 1843.938 & [1,1,1,1,2] & 1 \\
Prime Video Live & 8.266 & 20.444 & 106.283 & 853.313 & 1198.881 & [2,1,1,1,2] & 2 \\
Prime Video LongVideo & 3.761 & 17.660 & 199.026 & 2232.049 & 1851.821 & [1,1,1,1,2] & 1 \\
TikTok Browse & 1.161 & 11.499 & 136.573 & 480.831 & 796.071 & [1,1,1,1,2] & 1 \\
TikTok Live & 1.049 & 3.501 & 22.653 & 95.712 & 46.384 & [1,0,0,0,1] & 4 \\
TikTok ShortVideo & 1.211 & 12.772 & 309.290 & 1380.829 & 1983.031 & [1,1,1,1,2] & 1 \\
YouTube Browse & 2.029 & 17.398 & 71.809 & 552.994 & 734.962 & [1,1,1,1,2] & 1 \\
YouTube Live & 1.287 & 15.469 & 109.855 & 1013.868 & 928.634 & [1,1,1,1,2] & 1 \\
YouTube LongVideo & 0.576 & 24.026 & 388.746 & 4124.126 & 4751.549 & [1,1,1,2,2] & 0 \\
YouTube ShortVideo & 2.221 & 41.676 & 159.914 & 2822.534 & 3488.171 & [1,1,1,1,2] & 1 \\
Zoom Audio & 0.056 & 1.170 & 23.013 & 70.206 & 14.184 & [0,0,0,0,0] & 3 \\
Zoom BiVideo & 2.341 & 1.848 & 6.158 & 60.906 & 15.694 & [1,0,0,0,0] & 5 \\
Zoom DownVideo & 2.077 & 1.904 & 7.249 & 58.149 & 9.431 & [1,0,0,0,0] & 5 \\
Zoom UpVideo & 2.130 & 2.246 & 8.139 & 58.725 & 7.739 & [1,0,0,0,0] & 5 \\
\midrule
\textit{Metric Classes} & \textit{3 classes (0-2)} & \textit{2 classes (0-1)} & \textit{2 classes (0-1)} & \textit{3 classes (0-2)} & \textit{3 classes (0-2)} & & \\
\bottomrule
\end{tabular}
\end{table*}

\begin{figure*}[ht]
    \centering
    \subfigure[Baseline NTC.]{\includegraphics[width = 0.32\linewidth]{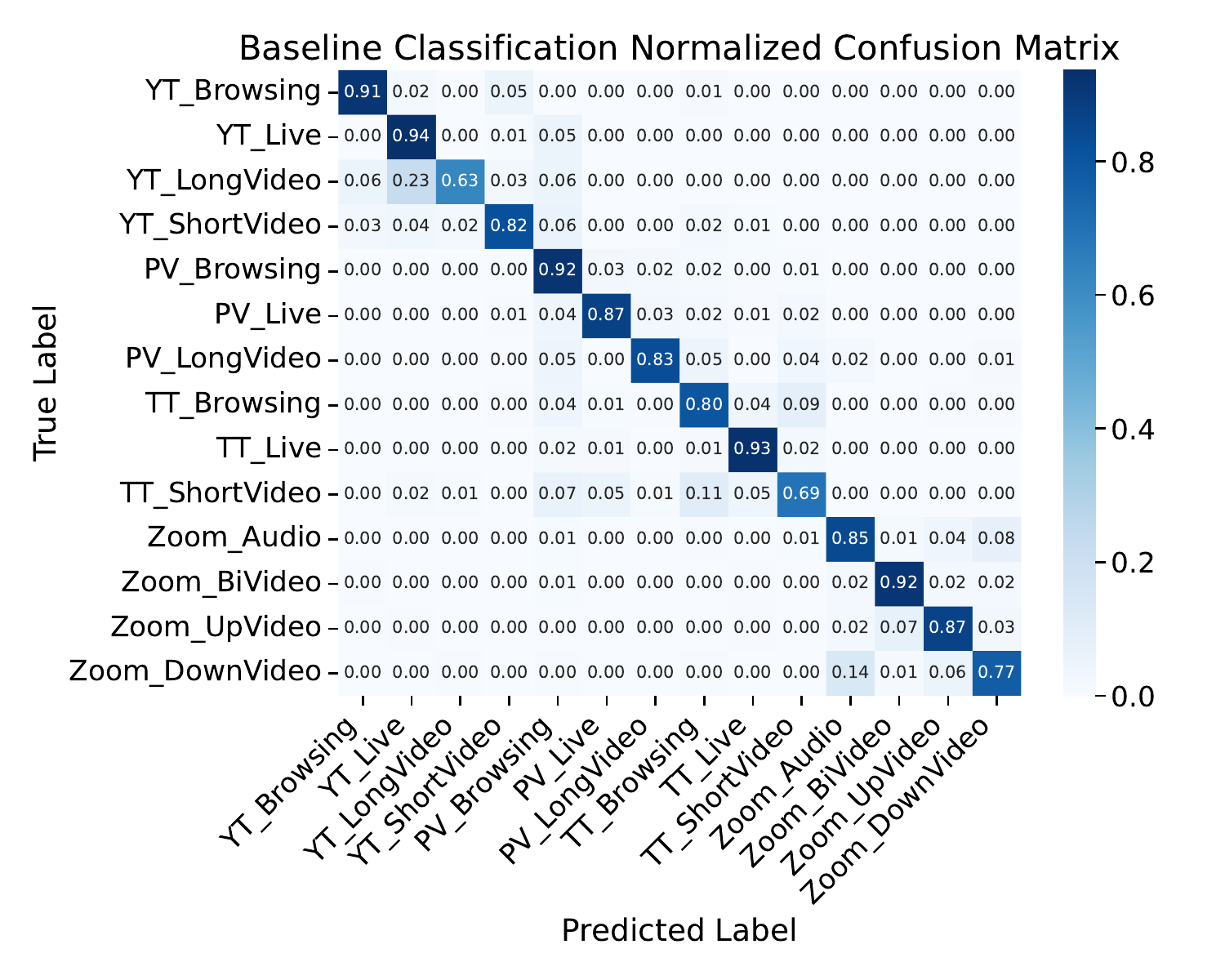}}\label{fig:BS-CM}%
    \subfigure[QoS-aware NTC.]{\includegraphics[width = 0.32\linewidth]{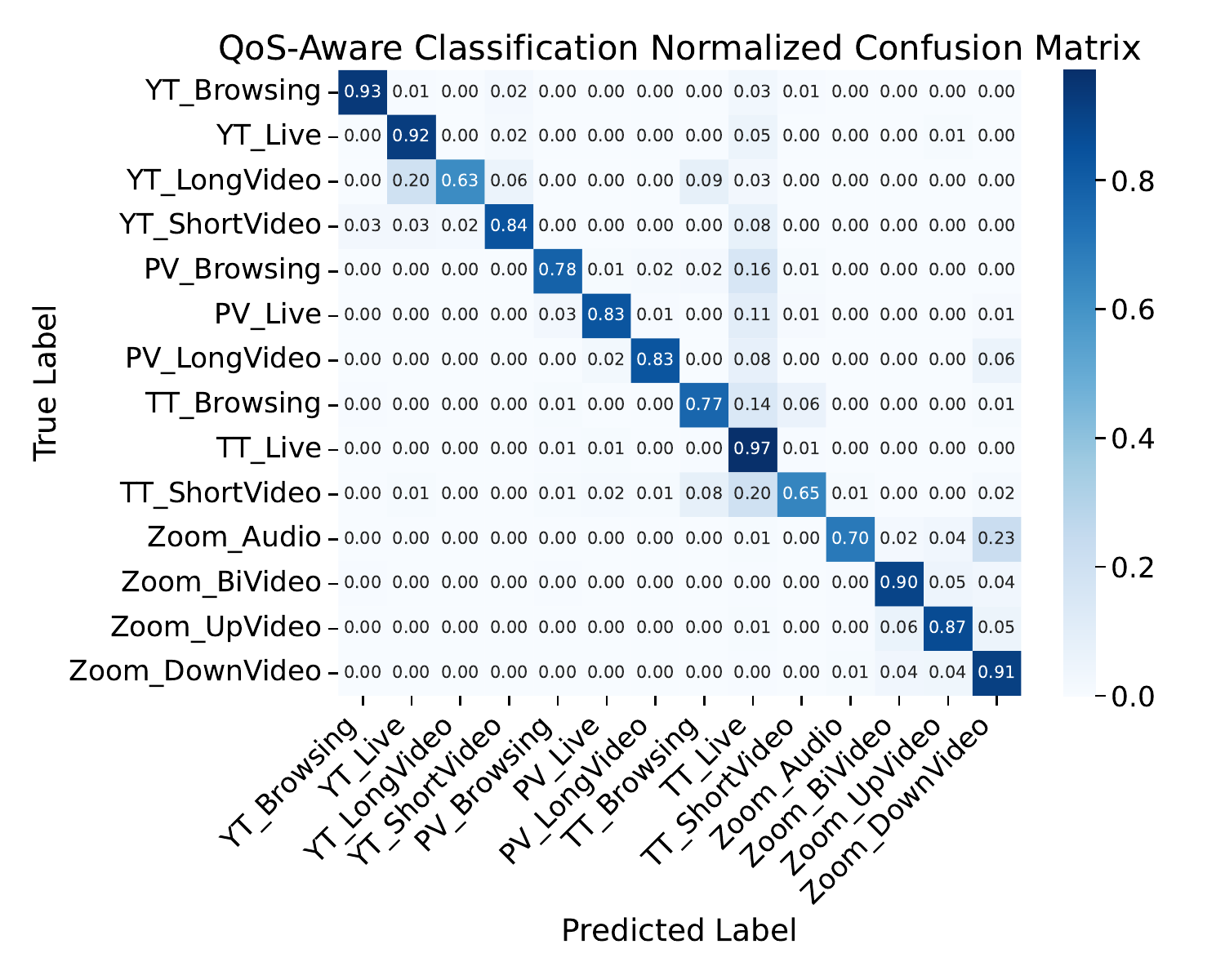}}\label{fig:QA-CM}
    \subfigure[Packet-level NTC~\cite{9882011}.]{\includegraphics[width = 0.32\linewidth]{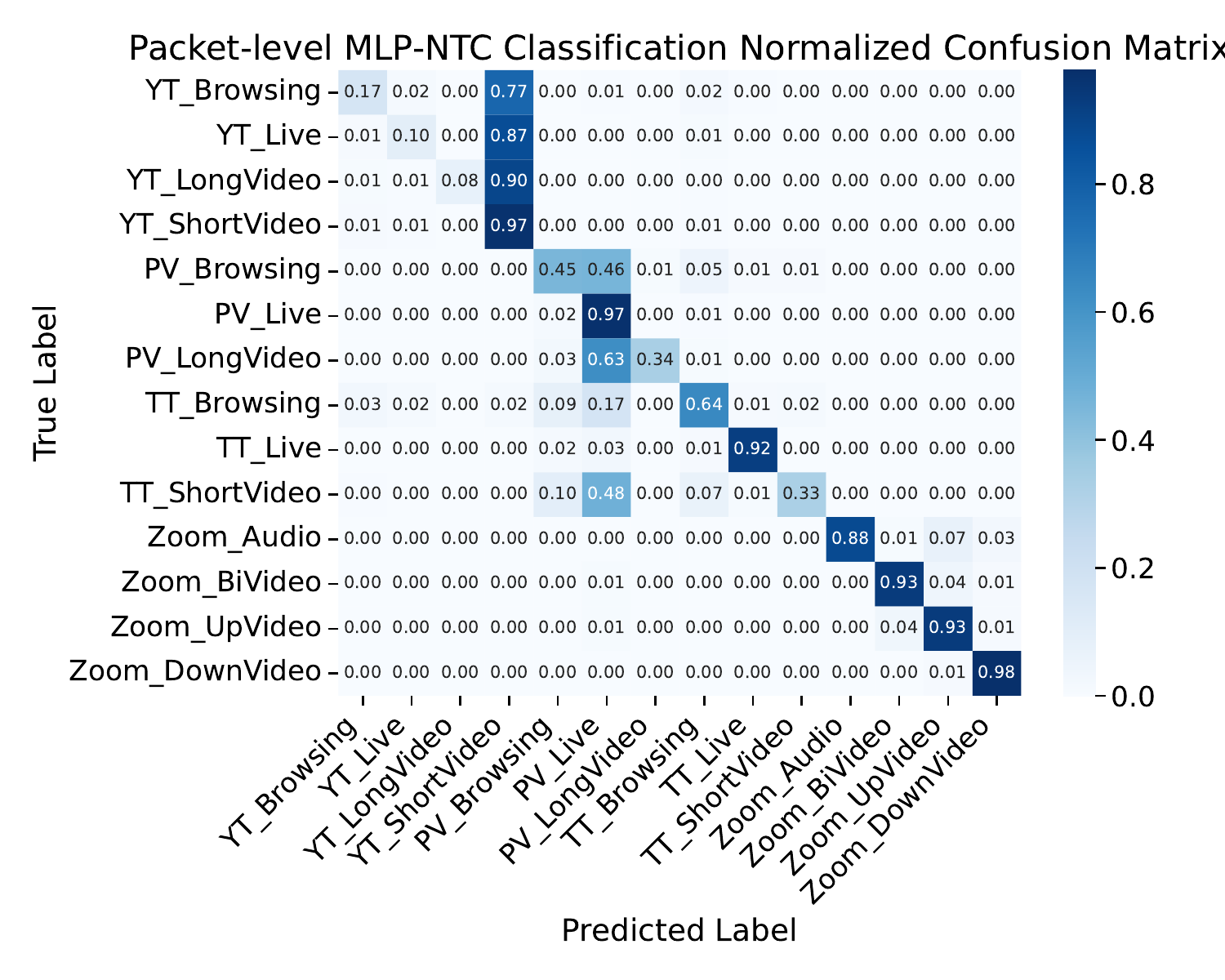}}\label{fig:Packet-CM}
\caption{Normalized confusion matrices for baseline NTC, proposed QoS-aware NTC, and an existing packet-level NTC. } \label{fig:NTC_comparsion}
\end{figure*}

\subsection{Performance Evaluation on QoS-Aware NTC}\label{sec:QoS_e}

QoS awareness is first extracted before implementing the QoS-aware NTC. Each of the fourteen application usage scenario is represented by a five-element class sequence, constructed by concatenating its class IDs across the five QoS metrics in the following order: [bandwidth, jitter stability, packet stability, burst frequency, burst stability], as detailed in Table~\ref{tab:qos_metrics}. Using the magnitude-based classification algorithm described in Sec.~\ref{sec:qos_alg}, the QoS awareness scores are derived for each application usage scenario. In this study, the magnitude threshold $X_{\text{thresh}}$ is set to 1.0, resulting in the following class distributions: bandwidth is divided into three classes (0,1,2), while jitter stability, packet stability, burst frequency, and burst stability are each divided into two classes (0,1). Based on identical class sequences, the algorithm identifies six distinct QoS awareness groups among the fourteen usages.

The largest group, Group~1, includes seven usage scenarios, including Prime Video (Browsing and Long-form Video), TikTok (Browsing and Short-form Video), and YouTube (Browsing, Live, and Short-form Video), all sharing the class sequence [1, 1, 1, 1, 2], indicative of moderate bandwidth and stability requirements. In contrast, Group~5 achieves the highest QoS awareness level (score 5), comprising three real-time application usage scenarios including Zoom video conferencing modes. These scenarios exhibit the class sequence [1, 0, 0, 0, 0], reflecting high stability demands and low tolerance for jitter and burst variability.

To rank the QoS groups by priority, a weighted scoring mechanism is applied using the following metric weights: bandwidth (30\%), jitter stability (20\%), packet stability (15\%), burst frequency (20\%), and burst stability (15\%). For example, Group~5 achieves the highest weighted score of 1.35 due to its optimal stability profile, while Group~0 receives the lowest score of 0.30, reflecting its relatively relaxed QoS requirements. This ranking framework enables priority-based QoS management, where traffic flows belonging to higher-scored groups are granted preferential treatment in network resource allocation. Such differentiation is critical for maintaining service quality in latency-sensitive and real-time applications.

\begin{table}[ht]
\centering
\vspace{+2mm}
\caption{Classification results comparison between baseline NTC and QoS-aware NTC.}
\label{tab:classification_results_comparison}
\footnotesize
\begin{tabular}
{@{}p{0.18\linewidth}|p{0.09\linewidth}p{0.09\linewidth}p{0.07\linewidth}|p{0.09\linewidth}p{0.09\linewidth}p{0.07\linewidth}@{}}
\toprule
\multirow{2}{*}{\textbf{Class}} & \multicolumn{3}{c|}{\textbf{Baseline NTC}} & \multicolumn{3}{c}{\textbf{QoS-aware NTC}} \\
\cmidrule(lr){2-4} \cmidrule(lr){5-7}
& \textbf{Precision} & \textbf{Recall} & \textbf{F1} & \textbf{Precision} & \textbf{Recall} & \textbf{F1} \\
\midrule
YT\_Browsing & 0.96 & 0.91 & 0.94 & 0.97 & 0.93 & 0.95 \\
YT\_Live & 0.85 & 0.94 & 0.89 & 0.89 & 0.92 & 0.91 \\
YT\_LongVideo & 0.85 & 0.63 & 0.72 & 0.92 & 0.63 & 0.75 \\
YT\_ShortVideo & 0.87 & 0.82 & 0.84 & 0.91 & 0.84 & 0.88 \\
PV\_Browsing & 0.79 & 0.92 & 0.85 & 0.94 & 0.78 & 0.85 \\
PV\_Live & 0.86 & 0.87 & 0.86 & 0.92 & 0.83 & 0.87 \\
PV\_LongVideo & 0.85 & 0.83 & 0.84 & 0.92 & 0.83 & 0.88 \\
TT\_Browsing & 0.82 & 0.80 & 0.81 & 0.89 & 0.77 & 0.82 \\
TT\_Live & 0.95 & 0.93 & 0.94 & 0.75 & 0.97 & 0.85 \\
TT\_ShortVideo & 0.70 & 0.69 & 0.70 & 0.79 & 0.65 & 0.71 \\
Zoom\_Audio & 0.84 & 0.85 & 0.84 & 0.97 & 0.70 & 0.81 \\
Zoom\_Bi & 0.90 & 0.92 & 0.91 & 0.88 & 0.90 & 0.89 \\
Zoom\_Up & 0.88 & 0.87 & 0.87 & 0.87 & 0.87 & 0.87 \\
Zoom\_Down & 0.83 & 0.77 & 0.80 & 0.68 & 0.91 & 0.78 \\
\midrule
\textbf{Accuracy} & & & \textbf{0.86} & & & \textbf{0.85} \\
\textbf{Mac. Avg} & \textbf{0.85} & \textbf{0.84} & \textbf{0.84} & \textbf{0.88} & \textbf{0.82} & \textbf{0.84} \\
\textbf{Wtd. Avg} & \textbf{0.86} & \textbf{0.86} & \textbf{0.86} & \textbf{0.86} & \textbf{0.85} & \textbf{0.85} \\
\bottomrule
\end{tabular}
\end{table}

The QoS awareness mechanism is then integrated into the proposed NTC framework. For comparison purposes, a standard MLP classifier that does not incorporate QoS awareness is used as the baseline. Before presenting the QoS performance metrics, we first evaluate the traditional classification accuracy. As illustrated in Fig.~\ref{fig:NTC_comparsion}, both the baseline classifier and the QoS-aware classifier achieve high accuracy across various application usage scenarios. A closer examination of Table~\ref{tab:classification_results_comparison} shows that, although the overall accuracy of the QoS-aware model is slightly lower than that of the baseline, the difference is marginal. In fact, the average weighted accuracy remains the same for both models. Furthermore, the QoS-aware approach demonstrates improved classification accuracy for certain specific usage types, highlighting its ability to capture nuanced in-app behaviors. In contrast, a state-of-the-art packet-level NTC method~\cite{9882011} achieves only 72.7\% accuracy across all usage scenarios. This lower performance is largely due to its tendency to misclassify usage scenarios that originate from the same application, which can mislead QoS provisioning.

\begin{figure}[ht!]
    \centering
    \subfigure[QoS performance with baseline NTC.]{\includegraphics[width = 0.9\linewidth]{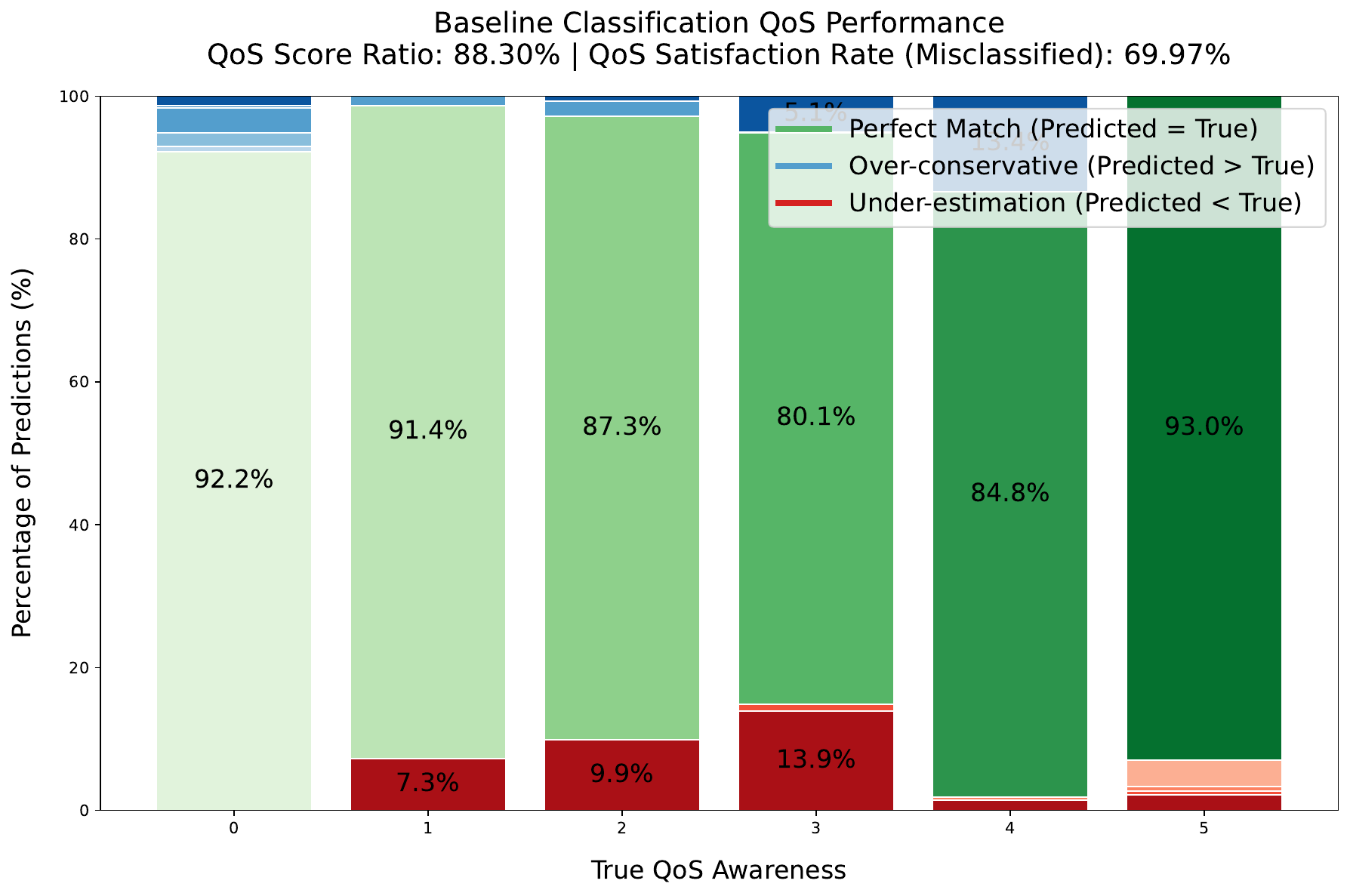}}\label{fig:QoS_org}\\
    \subfigure[QoS performance with the proposed fine-grained NTC.]{\includegraphics[width = 0.9\linewidth]{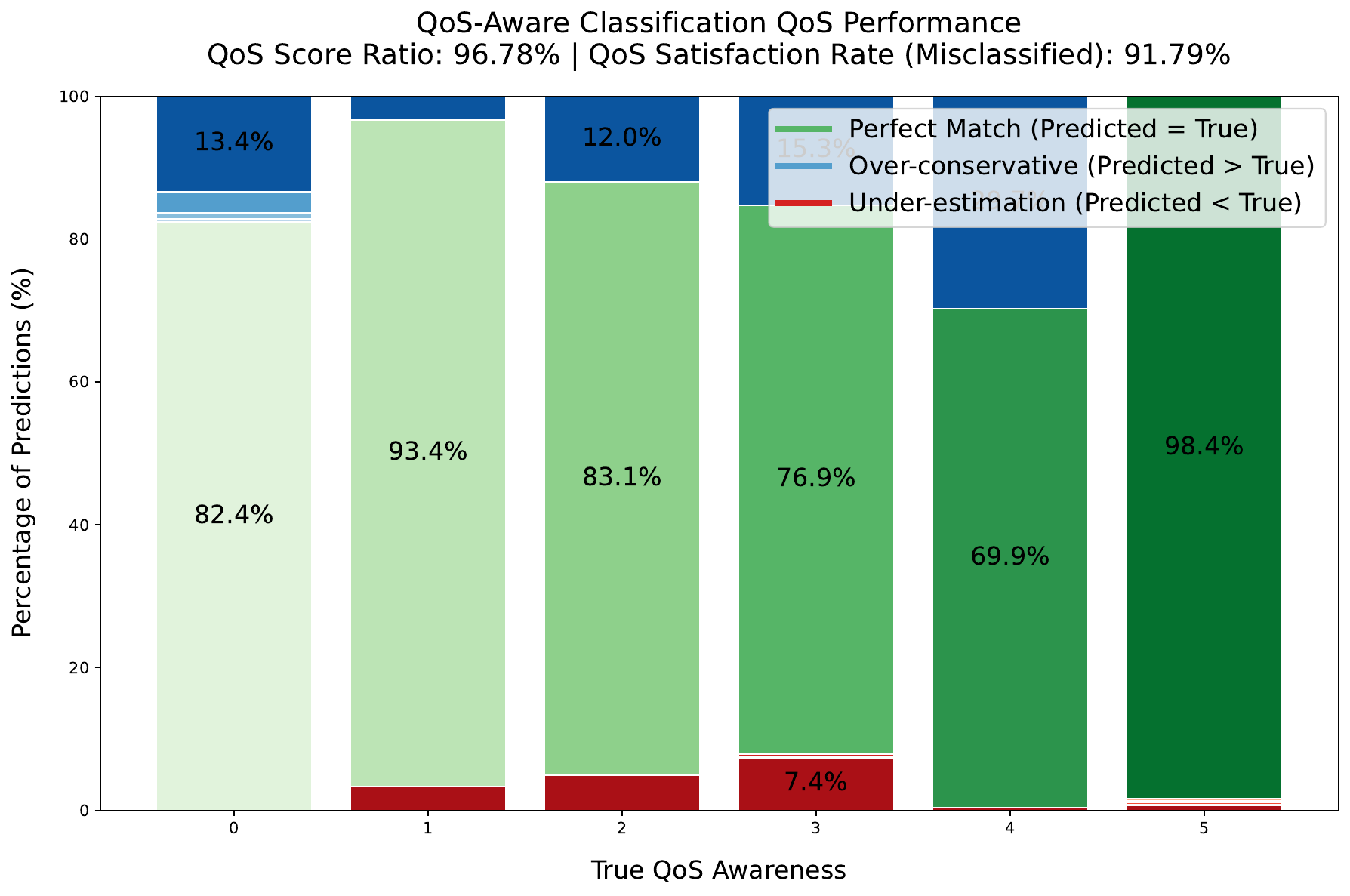}}\label{fig:QoS_aware}
\caption{QoS performance comparison between baseline and proposed fine-grained NTC models.} \label{fig:QoS_comparsion}    
\end{figure}

We then demonstrate the improved QoS performance from the QoS-aware NTC. As shown in Fig.~\ref{fig:QoS_comparsion}, the QoS-aware model achieves a significantly higher QoS score ratio of 96.78 compared to the baseline's 88.30, representing an improvement of 8.48 points. Additionally, for misclassified samples, the QoS-aware model achieves a satisfaction rate of 91.79\% compared to the baseline's 69.97\%. The performance distribution figures reveal that while the baseline model may achieve higher overall classification accuracy, its misclassifications often fail to meet QoS requirements, as evidenced by a larger proportion of under-provisioned cases. In contrast, the QoS-aware model shows a substantially smaller proportion of misclassifications that fail to satisfy QoS level requirements, ensuring better service quality for applications. However, this conservative approach may lead to over-provisioning in some cases, potentially resulting in resource wastage as the model tends to assign higher QoS levels to avoid service degradation.

\subsection{More Discussion}

A fundamental design decision in the proposed framework is to classify usage patterns first, rather than directly predicting QoS awareness levels as target labels. By decoupling these two stages, the framework gains greater stability, flexibility, and interpretability. Usage classification remains consistent and reusable across different network environments, while QoS policies can be dynamically adapted based on evolving service requirements or resource constraints.
The experimental results yield several key insights into the effectiveness of the proposed QoS-aware hierarchical GNN framework for fine-grained network traffic classification. Notably, the three-level hierarchical graph structure successfully addresses the limitations of single-scale approaches by enabling the model to learn both local and contextual features. The evaluation results proves its effectiveness in capturing multi-scale temporal dependencies, which ensure accurate and robust classification in fine-grained usage scenarios. Meanwhile, the QoS awareness is not necessarily obtained with a trade-off from the traditional classification accuracy. In fact, the evalution results demonstrated a slightly improved classification accuracy. It is because the QoS awareness impacts more on the uncertain classification results, which are highly likely to be misclassified by a normal NTC. The QoS awareness alters the final output, which may lead to a correct output. Meanwhile, the QoS-aware model significantly improves the QoS Experience Score (96.78 vs. 88.30). This improvement reflects the model’s conservative bias toward over-provisioning, which is preferable in practical network management scenarios where under-provisioning can lead to service degradation, whereas temporary over-allocation is generally more tolerable. 

Despite these strengths, several limitations warrant consideration. The current evaluation focuses on four major applications, which may limit generalizability to broader traffic domains. Additionally, the conservative QoS bias, while beneficial for service assurance, may lead to inefficient resource utilization in bandwidth-constrained environments. These observations suggest promising directions for future work, including dynamic adjustment of QoS weighting strategies and expansion to a wider range of application types and network conditions.

\section{Conclusion and Future Works}\label{sec:conclusion}

This paper presented a hierarchical GNN framework designed for fine-grained, QoS-aware network traffic classification. By integrating multi-scale graph modeling with a five-attribute  QoS awareness assignment algorithm, the proposed framework enables accurate differentiation of in-app usage patterns while maintaining a strong focus on service quality. Experimental results demonstrate that the developed GNN framework outperforms a state-of-the-art NTC method in fine-grained service-level application identification, achieving an accuracy of 86\% compared to 72.9\%. Furthermore, the inclusion of QoS-aware adjustment within the overall GNN framework does not negatively impact the overall classification accuracy. On the contrary, it significantly enhances the QoS experience, with a notable improvement in the QoS score (96.78 vs. 88.30) and the QoS satisfaction rate (91.79\% vs. 69.97\%). This improvement is particularly valuable in real-world network environments, where preserving service quality is essential. To further improve the adaptability and efficiency of the framework, future work will focus on dynamic QoS bias adjustment based on real-time network conditions. Additionally, the framework will be extended to support a wider range of application types and deployment scenarios.

\renewcommand\refname{Reference}
\bibliographystyle{IEEEtran}
\bibliography{Reference}

@misc{pcapdroid,
  author = {Emanuele Fusillo},
  title = {PCAPdroid: No-root network monitor, firewall and PCAP dumper for Android},
  year = {2020},
  howpublished = {\url{https://github.com/emanuele-f/PCAPdroid}},
  note = {Accessed: 2025-04-29}
}

@article{DBLP:journals/corr/abs-2105-14491,
  author       = {Shaked Brody and
                  Uri Alon and
                  Eran Yahav},
  title        = {How Attentive are Graph Attention Networks?},
  journal      = {CoRR},
  volume       = {abs/2105.14491},
  year         = {2021},
  url          = {https://arxiv.org/abs/2105.14491},
  eprinttype    = {arXiv},
  eprint       = {2105.14491},
  bibsource    = {dblp computer science bibliography, https://dblp.org}
}

@INPROCEEDINGS{10190520,
  author={Apruzzese, Giovanni and Laskov, Pavel and Schneider, Johannes},
  booktitle={2023 IEEE 8th European Symposium on Security and Privacy (EuroS\&P)}, 
  title={SoK: Pragmatic Assessment of Machine Learning for Network Intrusion Detection}, 
  year={2023},
  volume={},
  number={},
  pages={592-614},
  doi={10.1109/EuroSP57164.2023.00042}}

@INPROCEEDINGS{10179289,
  author={Mathews, Nate and Holland, James K and Oh, Se Eun and Rahman, Mohammad Saidur and Hopper, Nicholas and Wright, Matthew},
  booktitle={2023 IEEE Symposium on Security and Privacy (SP)}, 
  title={SoK: A Critical Evaluation of Efficient Website Fingerprinting Defenses}, 
  year={2023},
  volume={},
  number={},
  pages={969-986},
  doi={10.1109/SP46215.2023.10179289}}

@INPROCEEDINGS{9488690,
  author={Diallo, Alec F. and Patras, Paul},
  booktitle={IEEE INFOCOM 2021 - IEEE Conference on Computer Communications}, 
  title={Adaptive Clustering-based Malicious Traffic Classification at the Network Edge}, 
  year={2021},
  volume={},
  number={},
  pages={1-10},
  doi={10.1109/INFOCOM42981.2021.9488690}}

@inproceedings{yang2021cade,
  title={$\{$CADE$\}$: Detecting and explaining concept drift samples for security applications},
  author={Yang, Limin and Guo, Wenbo and Hao, Qingying and Ciptadi, Arridhana and Ahmadzadeh, Ali and Xing, Xinyu and Wang, Gang},
  booktitle={30th USENIX Security Symposium (USENIX Security 21)},
  pages={2327--2344},
  year={2021}
}

@ARTICLE{10870109,
  author={Abbasi, Mahmoud and López Flórez, Sebastián and Shahraki, Amin and Taherkordi, Amir and Prieto, Javier and Corchado, Juan M.},
  journal={IEEE Access}, 
  title={Class Imbalance in Network Traffic Classification: An Adaptive Weight Ensemble-of-Ensemble Learning Method}, 
  year={2025},
  volume={13},
  number={},
  pages={26171-26192},
  doi={10.1109/ACCESS.2025.3538170}}

@inproceedings{wang2023bars,
  title={BARS: Local robustness certification for deep learning based traffic analysis systems.},
  author={Wang, Kai and Wang, Zhiliang and Han, Dongqi and Chen, Wenqi and Yang, Jiahai and Shi, Xingang and Yin, Xia},
  booktitle={NDSS},
  year={2023}
}

@article{zhao2024towards,
  title={: Towards Fine-Grained Unknown Class Detection Against the Open-Set Attack Spectrum With Variable Legitimate Traffic},
  author={Zhao, Ziming and Li, Zhaoxuan and Xie, Xiaofei and Yu, Jiongchi and Zhang, Fan and Zhang, Rui and Chen, Binbin and Luo, Xiangyang and Hu, Ming and Ma, Wenrui},
  journal={IEEE/ACM Transactions on Networking},
  year={2024},
  publisher={IEEE}
}

@ARTICLE{5936162,
  author={Park, Byungchul and Hong, James Won-Ki and Won, Young J.},
  journal={IEEE Communications Magazine}, 
  title={Toward fine-grained traffic classification}, 
  year={2011},
  volume={49},
  number={7},
  pages={104-111},
  doi={10.1109/MCOM.2011.5936162},
  ISSN={1558-1896},
  month={July},}

@INPROCEEDINGS{7020869,
  author={Lin, Po-Ching and Chen, Shian-Yi and Lin, Chi-Hung},
  booktitle={2014 Australasian Telecommunication Networks and Applications Conference (ATNAC)}, 
  title={Towards fine-grained traffic classification for web applications}, 
  year={2014},
  volume={},
  number={},
  pages={28-33},
  doi={10.1109/ATNAC.2014.7020869}}

@ARTICLE{7377112,
  author={Fu, Yanjie and Xiong, Hui and Lu, Xinjiang and Yang, Jin and Chen, Can},
  journal={IEEE Transactions on Mobile Computing}, 
  title={Service Usage Classification with Encrypted Internet Traffic in Mobile Messaging Apps}, 
  year={2016},
  volume={15},
  number={11},
  pages={2851-2864},
  doi={10.1109/TMC.2016.2516020}}

@ARTICLE{9319399,
  author={Shen, Meng and Zhang, Jinpeng and Zhu, Liehuang and Xu, Ke and Du, Xiaojiang},
  journal={IEEE Transactions on Information Forensics and Security}, 
  title={Accurate Decentralized Application Identification via Encrypted Traffic Analysis Using Graph Neural Networks}, 
  year={2021},
  volume={16},
  number={},
  pages={2367-2380},
  doi={10.1109/TIFS.2021.3050608}}

@ARTICLE{9933044,
  author={Li, Wenhao and Zhang, Xiao-Yu and Bao, Huaifeng and Shi, Haichao and Wang, Qiang},
  journal={IEEE/ACM Transactions on Networking}, 
  title={ProGraph: Robust Network Traffic Identification With Graph Propagation}, 
  year={2023},
  volume={31},
  number={3},
  pages={1385-1399},
  doi={10.1109/TNET.2022.3216603}}

@ARTICLE{9882011,
  author={Zhang, Jielun and Li, Fuhao and Ye, Feng},
  journal={IEEE/ACM Transactions on Networking}, 
  title={Sustaining the High Performance of AI-Based Network Traffic Classification Models}, 
  year={2023},
  volume={31},
  number={2},
  pages={816-827},
  doi={10.1109/TNET.2022.3203227},
  ISSN={1558-2566},
  month={April},}

@inproceedings{zhao2023yet,
  title={Yet another traffic classifier: A masked autoencoder based traffic transformer with multi-level flow representation},
  author={Zhao, Ruijie and Zhan, Mingwei and Deng, Xianwen and Wang, Yanhao and Wang, Yijun and Gui, Guan and Xue, Zhi},
  booktitle={Proceedings of the AAAI Conference on Artificial Intelligence},
  volume={37},
  pages={5420--5427},
  year={2023}
}

@ARTICLE{5280678,
  author={Satyanarayanan, Mahadev and Bahl, Paramvir and Caceres, Ramon and Davies, Nigel},
  journal={IEEE Pervasive Computing}, 
  title={The Case for VM-Based Cloudlets in Mobile Computing}, 
  year={2009},
  volume={8},
  number={4},
  pages={14-23},
  doi={10.1109/MPRV.2009.82}}

@article{10.1145/3652595,
author = {Alhakamy, A’aeshah},
title = {Extended Reality (XR) Toward Building Immersive Solutions: The Key to Unlocking Industry 4.0},
year = {2024},
issue_date = {September 2024},
publisher = {Association for Computing Machinery},
address = {New York, NY, USA},
volume = {56},
number = {9},
issn = {0360-0300},
url = {https://doi.org/10.1145/3652595},
doi = {10.1145/3652595},
journal = {ACM Comput. Surv.},
month = apr,
articleno = {237},
numpages = {38}
}

@ARTICLE{8246845,
  author={Campolo, Claudia and Molinaro, Antonella and Iera, Antonio and Menichella, Francesco},
  journal={IEEE Wireless Communications}, 
  title={5G Network Slicing for Vehicle-to-Everything Services}, 
  year={2017},
  volume={24},
  number={6},
  pages={38-45},
  doi={10.1109/MWC.2017.1600408}}

@ARTICLE{8819994,
  author={Aceto, Giuseppe and Persico, Valerio and Pescapé, Antonio},
  journal={IEEE Communications Surveys \& Tutorials}, 
  title={A Survey on Information and Communication Technologies for Industry 4.0: State-of-the-Art, Taxonomies, Perspectives, and Challenges}, 
  year={2019},
  volume={21},
  number={4},
  pages={3467-3501},
  doi={10.1109/COMST.2019.2938259}}

@article{AZAB2024676,
title = {Network traffic classification: Techniques, datasets, and challenges},
journal = {Digital Communications and Networks},
volume = {10},
number = {3},
pages = {676-692},
year = {2024},
issn = {2352-8648},
doi = {https://doi.org/10.1016/j.dcan.2022.09.009},
url = {https://www.sciencedirect.com/science/article/pii/S2352864822001845},
author = {Ahmad Azab and Mahmoud Khasawneh and Saed Alrabaee and Kim-Kwang Raymond Choo and Maysa Sarsour}
}

@article{YU20181209,
title = {QoS-aware Traffic Classification Architecture Using Machine Learning and Deep Packet Inspection in SDNs},
journal = {Procedia Computer Science},
volume = {131},
pages = {1209-1216},
year = {2018},
note = {Recent Advancement in Information and Communication Technology:},
issn = {1877-0509},
doi = {https://doi.org/10.1016/j.procs.2018.04.331},
url = {https://www.sciencedirect.com/science/article/pii/S1877050918307129},
author = {Changhe Yu and Julong Lan and JiChao Xie and Yuxiang Hu}
}

@ARTICLE{10459131,
  author={Beshley, Mykola and Kryvinska, Natalia and Beshley, Halyna and Panchenko, Oleksiy and Medvetskyi, Mykhailo},
  journal={Journal of Communications and Networks}, 
  title={Traffic engineering and QoS/QoE supporting techniques for emerging service-oriented software-defined network}, 
  year={2024},
  volume={26},
  number={1},
  pages={99-114},
  doi={10.23919/JCN.2023.000065}}

@inproceedings{10.1145/3097983.3098049,
author = {Liu, Junming and Fu, Yanjie and Ming, Jingci and Ren, Yong and Sun, Leilei and Xiong, Hui},
title = {Effective and Real-time In-App Activity Analysis in Encrypted Internet Traffic Streams},
year = {2017},
isbn = {9781450348874},
publisher = {Association for Computing Machinery},
address = {New York, NY, USA},
url = {https://doi.org/10.1145/3097983.3098049},
doi = {10.1145/3097983.3098049},
booktitle = {Proceedings of the 23rd ACM SIGKDD International Conference on Knowledge Discovery and Data Mining},
pages = {335–344},
numpages = {10},
location = {Halifax, NS, Canada},
series = {KDD '17}
}

@ARTICLE{9791420,
  author={Sheikh, Muhammad Sameer and Peng, Yinqiao},
  journal={IEEE Access}, 
  title={Procedures, Criteria, and Machine Learning Techniques for Network Traffic Classification: A Survey}, 
  year={2022},
  volume={10},
  number={},
  pages={61135-61158},
  doi={10.1109/ACCESS.2022.3181135}}

@ARTICLE{9395707,
  author={Shapira, Tal and Shavitt, Yuval},
  journal={IEEE Transactions on Network and Service Management}, 
  title={FlowPic: A Generic Representation for Encrypted Traffic Classification and Applications Identification}, 
  year={2021},
  volume={18},
  number={2},
  pages={1218-1232},
  doi={10.1109/TNSM.2021.3071441}}

@article{10.1145/3457904,
author = {Papadogiannaki, Eva and Ioannidis, Sotiris},
title = {A Survey on Encrypted Network Traffic Analysis Applications, Techniques, and Countermeasures},
year = {2021},
issue_date = {July 2022},
publisher = {Association for Computing Machinery},
address = {New York, NY, USA},
volume = {54},
number = {6},
issn = {0360-0300},
url = {https://doi.org/10.1145/3457904},
doi = {10.1145/3457904},
journal = {ACM Comput. Surv.},
month = jul,
articleno = {123},
numpages = {35}
}

@inproceedings{10.1145/3485832.3485925,
author = {Pham, Thai-Dien and Ho, Thien-Lac and Truong-Huu, Tram and Cao, Tien-Dung and Truong, Hong-Linh},
title = {MAppGraph: Mobile-App Classification on Encrypted Network Traffic using Deep Graph Convolution Neural Networks},
year = {2021},
isbn = {9781450385794},
publisher = {Association for Computing Machinery},
address = {New York, NY, USA},
url = {https://doi.org/10.1145/3485832.3485925},
doi = {10.1145/3485832.3485925},
booktitle = {Proceedings of the 37th Annual Computer Security Applications Conference},
pages = {1025–1038},
numpages = {14},
location = {Virtual Event, USA},
series = {ACSAC '21}
}

@ARTICLE{9979671,
  author={Huoh, Ting-Li and Luo, Yan and Li, Peilong and Zhang, Tong},
  journal={IEEE Transactions on Network and Service Management}, 
  title={Flow-Based Encrypted Network Traffic Classification With Graph Neural Networks}, 
  year={2023},
  volume={20},
  number={2},
  pages={1224-1237},
  doi={10.1109/TNSM.2022.3227500}}

@ARTICLE{9841019,
  author={Aouedi, Ons and Piamrat, Kandaraj and Parrein, Benoît},
  journal={IEEE Transactions on Network and Service Management}, 
  title={Ensemble-Based Deep Learning Model for Network Traffic Classification}, 
  year={2022},
  volume={19},
  number={4},
  pages={4124-4135},
  doi={10.1109/TNSM.2022.3193748}}

@article{DUAN2023206,
title = {Network traffic anomaly detection method based on multi-scale residual classifier},
journal = {Computer Communications},
volume = {198},
pages = {206-216},
year = {2023},
issn = {0140-3664},
doi = {https://doi.org/10.1016/j.comcom.2022.10.024},
url = {https://www.sciencedirect.com/science/article/pii/S0140366422004121},
author = {Xueyuan Duan and Yu Fu and Kun Wang}
}

@article{MA2021102215,
title = {A novel model for anomaly detection in network traffic based on kernel support vector machine},
journal = {Computers \& Security},
volume = {104},
pages = {102215},
year = {2021},
issn = {0167-4048},
doi = {https://doi.org/10.1016/j.cose.2021.102215},
url = {https://www.sciencedirect.com/science/article/pii/S0167404821000390},
author = {Qian Ma and Cong Sun and Baojiang Cui and Xiaohui Jin}
}

@article{pranto2022performance,
  title={Performance of machine learning techniques in anomaly detection with basic feature selection strategy-a network intrusion detection system},
  author={Pranto, Md Badiuzzaman and Ratul, Md Hasibul Alam and Rahman, Md Mahidur and Diya, Ishrat Jahan and Zahir, Zunayeed-Bin},
  journal={J. Adv. Inf. Technol},
  volume={13},
  number={1},
  year={2022}
}

@ARTICLE{9566310,
  author={Shahraki, Amin and Abbasi, Mahmoud and Taherkordi, Amir and Jurcut, Anca Delia},
  journal={IEEE Transactions on Cognitive Communications and Networking}, 
  title={Active Learning for Network Traffic Classification: A Technical Study}, 
  year={2022},
  volume={8},
  number={1},
  pages={422-439},
  doi={10.1109/TCCN.2021.3119062}}

\end{document}